\documentclass
[showpacs,superscriptaddress,10pt,balancelastpage,nofootinbib,twocolumn]{revtex4}%
\usepackage[verbose,colorlinks,hyperindex,breaklinks=true,pdfusetitle,citecolor=blue]{hyperref}
\usepackage{amsfonts}
\usepackage{amsmath}
\usepackage{amssymb}
\usepackage{graphicx}

\begin{document}
\title{Delicate and robust dynamical recurrences of matter waves in driven optical crystals}
\author{Muhammad Ayub\footnote{ayubok@gmail.com}}
\affiliation{Department of Electronics, Quaid-i-Azam University, 45320, Islamabad, Pakistan.}
\affiliation{Theoretical Plasma Physics Division, PINSTECH, Nilore, 45650, Islamabad, 
Pakistan.}
\author{Farhan Saif\footnote{farhan@qau.edu.pk}}
\affiliation{Department of Electronics, Quaid-i-Azam University, 45320, Islamabad, Pakistan.}
\affiliation{Center for Applied Physics and Mathematics, National University 
of Science and Technology, \\ Islamabad, Pakistan.}
\begin{abstract}
We study dynamical recurrences of a Bose-Einstein condensate in an optical crystal
subject to a periodic external driving force. The recurrence behavior of the condensate 
is analyzed as a function of time close to nonlinear resonances occurring in the classical counterpart.
Our mathematical formalism for the recurrence time scales is presented as {\it delicate recurrences,} which take 
place, for instance when the lattice is perturbed weakly and {\it robust recurrences,} which may manifest themselves
for a sufficiently strong external driving force.
The analysis not only is valid for a dilute condensate, but also is applicable for strongly interacting homogeneous 
condensate provided; the external modulation causes no significant change in 
density profile of the condensate. We explain parametric dependence of the dynamical recurrence times which can 
easily be realized in laboratory experiments. In addition, we find a good agreement between the obtained analytical 
results and numerical calculations.

\end{abstract}
\maketitle

\section{Introduction} 
Coherent control of matter waves is at the heart of many theoretical and experimental interests. 
Coherence of matter waves in optical crystals has evolved as an active research area
over last two decades and led to analyze fundamental characteristics of 
quantum mechanics \cite{BlochRMP2008}, quantum tunneling \cite{Grifoni1998}, 
Bloch oscillations \cite{AndersonScience1998}, quantum information \cite{Alber2001} and quantum chaos\cite{SaifPR,Raizen1999}. 
The fascinating developments in
this subject have wide applications from 
quantum metrology \cite{GiovannettiNatPho2011} to quantum corrals \cite{XiongPRA2010}. 
Time periodic modulations in 
matter wave optics have given birth both to hybrid 
nano-opto-mechanical systems \cite{KurnNature2008} and driven billiards \cite{EdsonPRE2010}. 
Hence, the theoretical and experimental advancements have led to explore coherent transport 
\cite{Ben1996,ZangPRA2010}, controlling 
ratchet effects \cite{HeimsothPRA2010}, dynamical localization \cite{HolthausPRA2009,Holthaus2010}, 
coherent destruction of 
tunneling \cite{Grossmann}, photon assisted tunneling 
\cite{WeissPRA2009}, entanglement \cite{CerffieldPRL2007}, precision measurement of gravitational 
acceleration \cite{PoliPRL2011} and extracting nearest-neighbor 
spin correlations in a fermionic Mott insulator \cite{greif2010probing} in driven optical crystals.
Furthermore, realization of strongly correlated phases \cite{GreinerNat2002}, coherent acceleration of 
matter wave \cite{SaifPRA2007} in 
driven optical crystals \cite{PottingPRA2001} have been studied in great details.

A quantum particle in its early evolution in a bounded system follows classical mechanics and reappears after a 
classical period 
following classical trajectory. Later, following wave mechanics, it spreads and collapses as a consequence of 
non-linearity of the energy spectrum. However, the discreteness of the quantum mechanics leads to its 
reconstruction at various longer time scales, which define quantum recurrence times, such as, quantum revival 
time and supper revival time.
We explain these recurrences of a quantum particle in a quantum chaotic system, which manifests complex dynamics 
in its classical counterpart. Due to their dependence on modulation effects, these dynamical recurrences are 
different 
from those taking place in an undriven system, and may be considered as a probe to study quantum chaos 
\cite{SaifPRE}.
Dynamical recurrences, in a bounded driven system, originate from the simultaneous excitation of discrete 
quasi-energy states 
\cite{Saif2005-a,SaifEPJD}.
In the present paper we explain matter waves dynamics in a phase modulated optical crystal and provide analytical 
relations 
for classical period, quantum revival and super-revival time scales. In addition, we explain the existence of 
delicate dynamical recurrences, for weak effective modulation, and robust dynamical 
recurrences for strong effective modulation of the periodically driven optical crystal. The classical counterpart 
of the dynamical system displays dominant regular dynamics and dominant stochastic dynamics, 
one after the other, as a function of increasing modulation amplitude \cite{Raizen1999}.   
Our analysis is valid for weakly interacting BECs, 
a situation that can be realized experimentally with Feshbach scattering resonances \cite{Vogels1997}. 
In addition, we suggest that the analysis is valid 
for strongly interacting homogeneous condensates as well, where, nonlinear term can be replaced by an effective 
potential provided the external modulation causes slight changes in density profile of the condensate as discussed in
Ref. \cite{ChoiNiu1999}. Hence, due to spatial and
temporal periodicity in the driven optical crystal, the quantum dynamics of the condensate inside a nonlinear resonance 
is effectively mapped on the Mathieu equation. Our analytical findings are confirmed by numerical results.

The paper is organized as follows: In Sec. \ref{sec:tds}, we model the effective Hamiltonian governing the time
evolution. Later, the quasi-energy 
spectrum for nonlinear resonances in periodically driven optical crystal is 
explained. We obtain mathematical relations for recurrence time scales for the two cases; delicate dynamical 
recurrences and
robust dynamical recurrences for matter waves in driven optical crystal in Sec.~\ref{sec:dcaol}. The analytical and 
numerical results are summarized in Sec. \ref{sec:NR}.

\section{The Model}
\label{sec:tds}
We consider a condensate which is loaded in the lower edge of the band corresponding to optical crystal 
to avoid dynamical and Landau instabilities \cite{BWuPRANJP2001}.
The dynamics of the matter wave in driven one dimensional optical crystal, strongly confined by 
radial trap is governed by the Hamiltonian \cite{StaliunasPRE2006,Holthaus2010},
\[
H=\frac{p^{2}}{2M}+\frac{\tilde V_o}{2}\cos[2k_{L}\{x-\Delta L\sin(\omega
_{m}t)\}]+g_{1D}|\tilde{\psi}|^2,
\]
where, $k_{L}$ is wave number, $\tilde V_o$ defines the potential depth of the lattice. Moreover, 
$\Delta L$ and $\omega_{m}$ are respectively, amplitude and
frequency of external drive, whereas, $M$ is the mass of an atom.
Furthermore, $g_{1D}=\hbar k_L\omega_{\perp}a_s$ 
defines the effective two body interaction
coefficient, $\omega_{\perp}$ is radial trap frequency and $a_s$ is inter-atomic 
s-wave scattering length. 

The unitary transformation\footnote{The unitary transformation is time
periodic and preserves the quasi-energy spectrum.} 
$\tilde{\psi}=\psi(x,t)\exp[\frac{i}{\hbar}\{\omega_{m}M\Delta L\cos(\omega_{m}t)%
x+\beta(t)\}],$ (where, $\beta(t)=\frac{\omega_{m}^{2}\Delta L^{2}M}%
{4}[\frac{\sin(2\omega_{m}t)}{2\omega_{m}}+t]$), for a frame co-moving with the
lattice, modifies the Hamiltonian as
\begin{equation}
H=\frac{p^{2}}{2M}+\frac{\tilde V_o}{2}\cos2k_{L}x-Fx\sin\omega_{m}t+g_{1D}|\psi|^2.
\label{ModHam}%
\end{equation}
Here, $F=M\Delta L\omega_{m}^{2}$ is amplitude of inertial force emerging in
the oscillating frame. To examine the dynamics of cold atoms in
driven optical crystals numerically, Hamiltonian (\ref{ModHam}) is expressed in
dimensionless quantities. We scale the quantities so that $z=k_{L}x,~\tau=\omega_{m}t$, 
$\psi=\frac{\psi}{\sqrt{n_0}}$, where, $\sqrt{n_0}$ is average density of a condensate. 
Multiplying the
Schr\"{o}dinger wave equation by $\frac{2\omega_{r}}{\hbar\omega_{m}^{2}},$
where, $\omega_{r}=\frac{\hbar k_{L}^{2}}{2M}$ is single\ photon recoil
frequency, we get dimensionless Hamiltonian, viz
\begin{equation}
\tilde{H}=-\frac{k^{\hspace{-2.1mm}-2}}{2}\frac{\partial^{2}}{\partial z^{2}%
}+\frac{V_o}{2}\cos2z+\lambda z\sin\tau +G|\psi|^2.\label{tdse}
\end{equation}
Here, $G=\frac{g_{1D}n_0 k^{\hspace{-2.1mm}-}}{\hbar \omega_m},$ 
and $\lambda=\frac{F~d_L~k^{\hspace{-2.1mm}-}}{\hbar\omega_m}=k_{L}\Delta L$ is scaled modulation amplitude. 
Also $V_o=\frac{\tilde V_o k^{\hspace{-2.1mm}-}%
}{2\hbar\omega_{m}},$ $\tau$ is scaled time in the units of modulation
frequency, $\omega_{m},$ and $d_L$ defines the periodicity of optical crystal. 
In this case, the rescaled Planck's constant is $k^{\hspace{-2.1mm}-}=\frac{2\omega_{r}}{\omega_{m}}.$

For the dynamics of weakly interacting BECs in optical crystal  
the interaction term can safely be neglected~\cite{HolthausPRA2009}, which is a situation 
achievable in the present day experiments with the advent of Feshbach scattering resonances.
In addition, the effective potential $\frac{V_o}{2}\cos 2z + G |\psi|^2$ seen by each atom 
may as well be written~\cite{ChoiNiu1999} as,
\[
V_{eff}=\frac{\acute{V}}{2}\cos(2z)+ const,
\]
where, $\acute{V}=\frac{V_o}{1+4G}$. 

The analytical result, calculated using perturbation theory~\cite{ChoiNiu1999}, 
is valid as long as the condensate density is nearly uniform, i.e. , $\acute{V} << 1$ which
describes either a weak external potential $V_o$ or a strong atomic interaction $G$.
The condition is experimentally confirmed for one dimensional potential \cite{MorschPRL2001}.
The effective dynamics of matter wave in this regime is governed by the Hamiltonian
\begin{equation}
\tilde{H}=-\frac{k^{\hspace{-2.1mm}-2}}{2}\frac{\partial^{2}}{\partial z^{2}%
}+\frac{\acute{V}}{2}\cos2z+\lambda z\sin\tau.\label{effctHal}
\end{equation}
Here and onward $\acute{V}$ 
is expressed in the units of recoil energy $E_r=\hbar\omega_r$.

\begin{figure}[h]
 \includegraphics[scale=0.85]{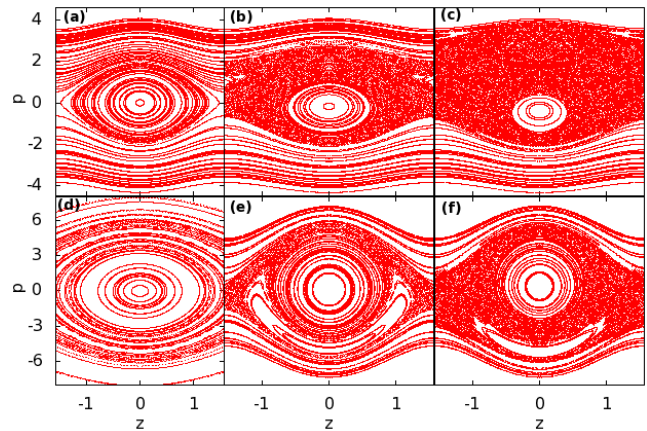}
  \caption{Poincare phase space for different modulation
 strengths and lattice potentials. Upper row: $\lambda=0$, (a) $,0.5$, (b) $1.5$, (c) 
and $\acute{V}=2$;
 Lower row: $\lambda=0$, (d) $0.5$, (e) $1.5$, (f) and $\acute{V}=16$.}%
 \label{fig:FigPhasespace}%
 \end{figure}

In order to understand the classical dynamics of the system, we plot the
Poincar\'e surface of section for shallow 
optical crystal ($\acute{V}=2$) with modulation strengths $\lambda=0$, $0.5$, $1$ and 
for deep lattice 
($\acute{V}=16$) with modulation strengths $\lambda=0$, $0.5$, $1.5$ as shown in 
Fig. \ref{fig:FigPhasespace}. From phase space plot, we see
the appearance of 1:1 resonance for $\lambda>0$. This resonance emerges when
the time period of external force matches with period of unperturbed system. One
effect of an external modulation is the development of stochastic region near the
separatrix. As modulation is increased, while the frequency is fixed, the size
of stochastic region increases at the cost of regular region.
 
In order to solve the time dependent Schr\"{o}dinger equation near the nonlinear resonances of 
the driven quantum system corresponding to the
Hamiltonian~(\ref{tdse}), we~\cite{Saif2002,Flatte,GPBerman1977}, provide an ansatz as 
\begin{equation}
|\psi(t)\rangle=\sum_{n}C_{n}(t)|n\rangle\exp[-i\{E_{\bar{n}}+(n-\bar{n}%
)\frac{k^{\hspace{-2.1mm}-}}{N}\}\frac{t}{k^{\hspace{-2.1mm}-}}], \label{psit}%
\end{equation}
where, $E_{\bar{n}}$ is the mean energy, $C_{n}(t)$ is time dependent
probability amplitude, $\bar{n}$ is mean quantum number, $|n\rangle$ are
eigen states of undriven system and $N$ is resonance number.

The Floquet energy spectrum for nonlinear resonances is given as \cite{Ayub2010,Holthaus1991},
\begin{equation}
\mathcal{E}_{\mu,\nu}=\left[ \frac{N^{2}k^{\hspace{-2.1mm}-2}\zeta}{8}a_{\nu
}(\mu(j),q)+k^{\hspace{-2.1mm}-}\tilde{\alpha}{j} \right] \,\, \mathrm{mod}
\,\,k^{\hspace{-2.1mm}-}\omega_m, \label{ener1}%
\end{equation}
where, $a_{\nu}(\mu(j),q)$ is Mathieu characteristics parameter and $q=\frac{\lambda}{\beta_0}$
is the effective modulation. Here, $\beta_0=\frac{N^{2}k^{\hspace{-2.1mm}-2}\zeta}{4V}$, and
$V=<n|z|n\pm 1>$ are matrix elements and are considered constant at potential minima. 
The solutions are defined by $\pi$-periodic Floquet states,
\textit{i. e.} $P_{\mu}(\theta)=P_{\mu}(\theta+ \pi)$, and $\mu$ is the characteristic exponent. In
order to have $\pi$-periodic solutions, we
require $\mu$ to be defined as $\mu=\mu(j)=2j/N$, where, $j=0,1,2,....,N-1$.

The allowed values of $\mu(j)$ can exist as a characteristic exponent of
solution to the Mathieu equation for discrete $\nu$ (which takes integer
values) only for certain value $a_{\nu}(\mu(j),q)$, when $q$ is fixed.
The index $\nu$ takes the definition $\nu=\frac{2(n-\bar{n})}{N}$. Moreover,
$\tilde{\alpha}$ defines the winding number.
For driven optical potential, keeping the periodicity of Floquet
solutions \cite{Saif2002}, we take only even values of the index
\cite{AbramowitzStegun}. Hence, the re-scaled Mathieu characteristic exponent becomes
\begin{equation}
\nu=2(l+\beta), \text{ where, } \beta=\frac{N\omega-1}{N^{2}\zeta
k^{\hspace{-2.1mm}-}}.
\end{equation}

Comparing the coefficients of eigen energy of the undriven system and equation
of motion of probability amplitude in the absence of modulation, we get
$l=\frac{n-\bar{n}}{N},$ which is new band index for nonlinear resonance and at
the center of resonance $l=0$. 
Analytical expressions of $a_{\nu}(\mu(j),q)$ for two asymptotic cases i.e. for $q\lesssim1$ 
and $q\gg1$ are different \cite{AbramowitzStegun}. We discuss these two cases in Sec-\ref{sec:dcaol} separately. 

\section{Recurrence Times in a Driven Optical Lattice
\label{sec:dcaol}}
As recurrence time scales of driven optical crystal are expressed in terms of undriven
system, we present classical period, quantum
revival time and super revival time scales of matter wave in undriven optical crystal
\cite{AyubJRLR2009,Holthaus1997}.

In the absence of external forcing, the crystal potential replaced with effective potential in equation 
(\ref{ModHam}) is governed by the Hamiltonian 
\begin{equation}
 \tilde H=\frac{p^{2}}{2}+\frac{\acute{V}}{2}\cos2z.
\end{equation}
The corresponding time independent Schr\"odinger equation is eventually a Mathieu equation, where, $q_0=\frac{\acute{V}}{4},$  
is effective crystal potential scaled by recoil energy.
In shallow lattice potential limit, i.e. $q_o\lesssim1$, the classical frequency, $\omega=2\bar{n}\{1-\frac
{q_o^{2}}{2(\bar{n}^{2}-1)^{2}}\}$ and non-linearity, $\zeta=2+\frac{q_o^{2}}%
{2}\frac{3\bar{n}^{2}+1}{(\bar{n}^{2}-1)^{3}}$.
The classical period of the condensate is inversely proportional to the average 
classical frequency, $\omega$, and calculated~\cite{AyubJRLR2009} as  
\begin{equation} 
T_{0}^{(cl)}=\left\{1+\frac{q_o^{2}}{2(\bar{n}^{2}-1)^{2}}\right\}\frac{\pi}{\bar{n}}.
\end{equation} 

In the long time evolution the condensate spreads and observes a collapse as a result of destructive interference 
of constituent wavelets. At longer time the constructive interference overcomes and leads to reconstruction of the 
condensate at quantum revival time, that is inversely proportional to the nonlinearity of the spectrum 
present around the mean quantum number, 
$\bar n$, viz.
\begin{equation}
 T_{0}^{(rev)}=2\pi\left\{1-\frac
{q_o^{2}}{2}\frac{(3\bar{n}^{2}+1)}{(\bar{n}^{2}-1)^{3}}\right\}.
\end{equation}
Whereas, the super-revival time, obtained as,
\begin{equation}
 T_{0}^{(spr)}=\frac{\pi(\bar{n}^{2}-1)^{4}}{q_o^{2}\bar{n}(\bar{n}^{2}+1)},
\end{equation}
is inversely proportional to incremental change in nonlinearity with respect to quantum number $n$, at mean quantum 
number $\bar n$.

In deep optical potential limit,
i.e. $q_o>>1$, the classical frequency is $\omega=4(\sqrt{q_o}-\frac
{2\bar{n}+1}{8})$ and non-linearity is $\zeta=|-1-\frac{3(2\bar{n}+1)}{2^{4}\sqrt{q_o}%
}|$.
In this case, the
classical time period, quantum revival time and super revival time are, respectively, calculated as
\begin{equation}
 T_{0}^{(cl)}=\frac{\pi}{2\sqrt{q_o}}\left\{1+\frac
{s}{8\sqrt{q_o}}+\frac{3(s^{2}+1)}{2^{8}q_o}\right\},
\end{equation}
 \begin{equation}
 T_{0}^{(rev)}=4\pi\left(1-\frac{3s}{16q_o}\right),
\end{equation}
\begin{equation}
\text{and \ \  }T_{0}^{(spr)}=32\pi\sqrt{q_o},
\end{equation}   where, $s=2\bar{n}+1$.

In the following we explain the quantum recurrences of condensate in optical crystal in 
the presence of 
external periodic forcing. For the sake of clarity and simplicity, we classify our later discussion as 
delicate dynamical recurrences for $q\lesssim1$ and robust dynamical recurrences for $q>>1$.

\subsection{Delicate dynamical recurrences}
For weakly driven, shallow or deep, lattice ($q\lesssim1$) the recurrence time scales, $T_{\lambda}^{(cl)}$, 
$T_{\lambda}^{(rev)}$ and $T_{\lambda}^{(spr)}$ for primary resonance $N=1$ are calculated as
\cite{Ayub2010}
\begin{equation}
T_{\lambda}^{(cl)}=T_{0}^{(cl)}\left[ 1+\frac{q^{2}}{2}\frac{1}{\{4(l+\beta
)^{2}-1\}^{2}}\right] \Delta, \label{TshallowCl}%
\end{equation}%
\begin{equation}
T_{\lambda}^{(rev)}=T_{0}^{(rev)}\left[ 1-\frac{q^{2}}{2}\frac{12(l+\beta
)^{2}+1}{\{4(l+\beta)^{2}-1\}^{3}}\right],  \label{TshallowRev}%
\end{equation}%
\begin{equation}
\text{and \ \  }T_{\lambda}^{(spr)}=\frac{\pi\{4(l+\beta)^{2}-1\}^{4}}%
{2\zeta k^{\hspace{-2.1mm}-}q^{2}(l+\beta)\{4(l+\beta)^{2}+1\}},
\label{TshallowSup}%
\end{equation}
where, $\Delta=(1-\frac{1}{N\omega})^{-1}$.
\begin{figure}[t]
\begin{center}
\includegraphics[scale=0.88]{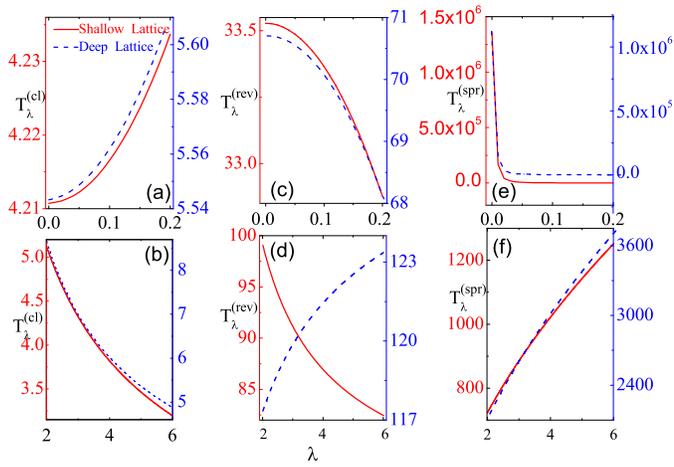}
\end{center}
\caption[r]{Left panel: Classical time period versus $\lambda=\frac{q}{\beta_0}$ for weak external
modulation (a) and for strong external modulation (b), where, 
$\beta_0=\frac{k^{\hspace{-2.1mm}-2}\zeta}{4V}$ for $N=1$. Middle panel: Quantum revival time 
versus $\lambda$ for weak external modulation (c) and for strong external modulation (d). 
Right panel: Super revival time versus $\lambda$ when external modulation is weak
(e) and for strong external modulation (f). In this figure, for deep
lattice $\acute {V}=16$,  for shallow lattice $\acute {V}=2$ and
$k^{\hspace{-2.1mm}-}=0.5$.}%
\label{fig:recure}%
\end{figure}
Here, time scales for weakly driven shallow lattice and weakly driven deep lattice 
are similar in structure, however, they differ due to different energy 
spectra corresponding to undriven crystal in the two cases which leads to different undriven time scales.

In order to observe the dynamics of a condensate inside a resonance, we
evolve a well localized Gaussian condensate in the driven optical
crystal.  Numerical results are obtained
by placing a condensate in a primary resonance with $N=1$. 

The classical period time, quantum revival time and super revival time of matter
waves in modulated optical crystal near nonlinear resonances versus scaled modulation 
$\lambda=\frac{q}{\beta_0}$ is
shown in Fig. \ref{fig:recure}. 
In each plot of the figure, left vertical axis shows the
time scale when shallow optical crystal is modulated and
right axis shows the time scale when deep optical crystal is modulated.
The upper row of Fig. \ref{fig:recure} represents the time scales for small $q$
values i.e. delicate dynamical recurrences as a function of
modulation $\lambda$, whereas, 
lower row represents the time scales when $q\gg1$, as a function of modulation
$\lambda$, i.e. robust dynamical recurrences. Here, left
column shows the results related to the classical periods. Quantum revival times 
are plotted in middle column, whereas, right column shows super revival times.

We note that  when optical
crystal is perturbed by weak periodic force, the classical period increases with
modulation, as given in equation (\ref{TshallowCl}). Classical period for weakly
driven shallow lattice potential changes slowly as compared to weakly driven
deep lattice potential as shown in Fig. \ref{fig:recure}a.
Quantum revival time in nonlinear resonances versus modulation is shown in
middle column of Fig. \ref{fig:recure}. For delicate dynamical recurrences
(Fig. \ref{fig:recure}c),  the
quantum revival time decreases as modulation increases. The behavior of
quantum revival time is given by equation
(\ref{TshallowRev}) for delicate
dynamical recurrence. For weakly driven 
shallow optical crystal or weakly driven deep lattice, qualitative and
quantitative behavior of revival time is almost similar. 
Classical period and quantum revival time for delicate dynamical
recurrences show good numerical and analytical resemblance for the
system with our previous work \cite{Saif2005-a,ShahidPLA,SaifEPJD}.

\begin{figure}[t]
 \includegraphics[scale=0.4]{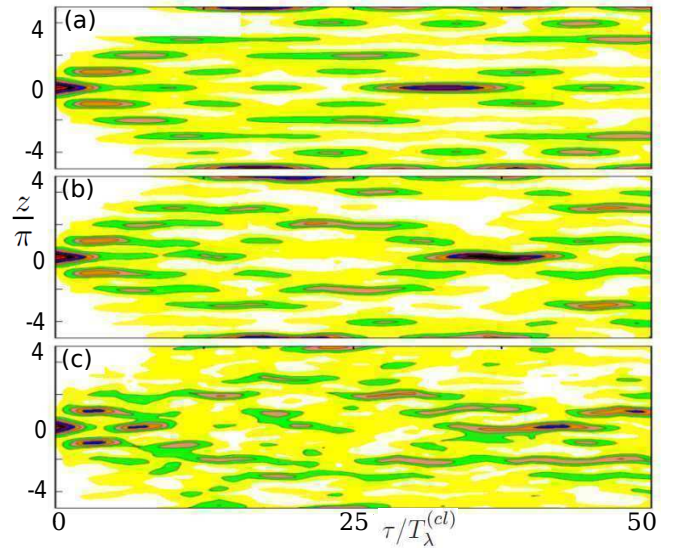}
 \caption{Spatio-temporal behavior of atomic condensate for $q=0 (\lambda=0)$, $\acute V=2$ (a)
$q=0.2\beta_0$, $\acute V=2$(b) and $q=0.2\beta_0$, $\acute V=0.3$ (c). Other
 parameters are $\Delta p=0.5$, $\beta_0=\frac{k^{\hspace{-2.1mm}-2}\zeta}{4V}$ and $k^{\hspace{-2.1mm}-}=1.$}%
 \label{fig:Spatiotemp2}%
 \end{figure}

Fig. \ref{fig:Spatiotemp2} shows spatio-temporal evolution of an initially well
localized condensate in a crystal potential well. Fig. \ref{fig:Spatiotemp2}a is spatio-temporal 
dynamics in undriven crystal, 
while Fig. \ref{fig:Spatiotemp2}b $\&$ Fig. \ref{fig:Spatiotemp2}c  present the case,
for external modulation $q=0.2\beta_0$ and different value of $\acute V$.  
Spatio-temporal evolution of the condensate
in optical crystal shows that condensate diffuses to the neighboring lattice sites 
by tunneling
and splits into smaller wavelets. Later, these wavelets
constructively interfere and condensate revival takes place. 
Revival time calculated numerically is the same as obtained from analytical results.
Keeping modulation constant as $q=0.2\beta_0$ but for different $\acute V$, which may be a consequence of 
varied atom-atom interaction, revival time is modified. 
We note that in the absence of interaction term, $G$, revival time changes with $\acute V$ and interference pattern
is similar. But with the introduction of interaction term 
not only revival time is modified due to change in $\acute V$, interference pattern is also modified
as seen in Fig. \ref{fig:Spatiotemp2}c.   

\subsection{Robust dynamical recurrences }
On the other hand, for strongly driven optical crystal, the effective modulation, $q\gg1$. 
Time scales for the atomic condensate for primary resonance with 
$N=1$ are given as \cite{Ayub2010}

\begin{equation}
T_{\lambda}^{(cl)}=\frac{2\pi}{k^{\hspace{-2.1mm}-}\zeta\left\{\sqrt{q}%
-\dfrac{4(l+\beta)+1}{8\ }\right\}}, \label{TdeepCl}%
\end{equation}%
\begin{equation}
T_{\lambda}^{(rev)}=\frac{8\pi}{k^{\hspace{-2.1mm}-}\zeta}\left[
1-\dfrac{3\{4(l+\beta)+1\}}{16\sqrt{q}}\right],  \label{TdeepRev}%
\end{equation}
\begin{equation}
\text{and \ \ \ \ \ \ \ \ \ \ \ \ \ \ }T_{\lambda}^{(spr)}=\frac{32\pi\ \sqrt{q}}{k^{\hspace{-2.1mm}-}\zeta}.
\label{TdeepSup}%
\end{equation}

In case of strongly driven crystal, when external modulation frequency is close to the
harmonic frequency, matrix elements, $V$ can be approximated by those of
harmonic oscillator and effective modulation, $q$, can be approximated as
$q\approx\frac{4\sqrt{n+1}\lambda}{q_o^{\frac{1}{4}}k^{\hspace{-2.1mm}%
-2}\zeta}$ \cite{Holthaus1997} . Under this approximation the time scales are%
\begin{equation}
T_{\lambda}^{(cl)}=\frac{16\pi q_o^{\frac{1}{8}} \text{/} \sqrt{\zeta}%
}{16(\bar{n}+1)^{\frac{1}{4}}\sqrt{\lambda}-\{4(l+\beta)+1\}q_o^{\frac{1}{8}%
}k^{\hspace{-2.1mm}-}\sqrt{\zeta}}, \label{TdeepClHar}%
\end{equation}
\[
T_{\lambda}^{(rev)}=\frac{8\pi}{k^{\hspace{-2.1mm}-}\zeta}\left[
1-\frac{3\{4(l+\beta)+1\}q_o^{\frac{1}{8}}k^{\hspace{-2.1mm}-}\sqrt{\zeta}%
}{32(\bar{n}+1)^{\frac{1}{4}}\sqrt{\lambda}}\right],
\]
\begin{equation}
\text{and \ \ \ \ \ \ }T_{\lambda}^{(spr)}=\frac{64\pi(\bar{n}+1)^{\frac{1}{4}}\sqrt{\lambda}%
}{k^{\hspace{-2.1mm}-2}\zeta^{\frac{3}{2}}q_o^{\frac{1}{8}}}.
\label{TdeepSprHar}%
\end{equation}

When lattice is strongly modulated by an external periodic force, the
classical period decreases as modulation increases. Classical period for
strongly driven optical crystal is given by equation (\ref{TdeepCl}). The
behavior of classical period for strongly driven lattice versus modulation is
qualitatively of the same order for both strongly driven shallow lattice and
strongly driven deep lattice as shown in Fig. \ref{fig:recure}b. The behavior
of classical period in strongly driven crystal case is understandable as
strong modulation influence more energy bands of undriven crystal to
follow the external frequency and near the center of nonlinear resonance the
energy spectrum is almost linear \cite{Ayub2010} with assumptions $q\gg1$ and $l$ is
small, i.e. condensate is placed near the center of resonance.

\begin{figure}[h]
 \includegraphics[scale=0.6]{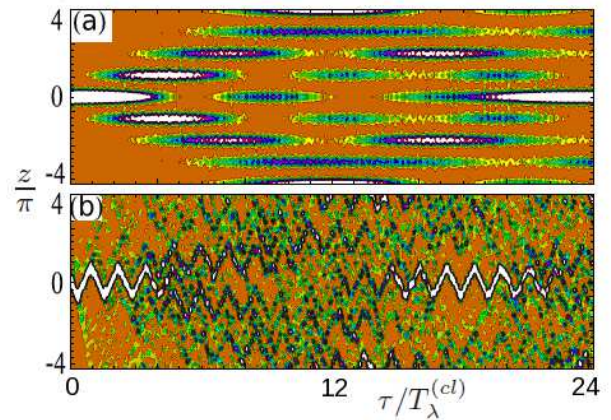}
  \caption{Spatiotemporal behavior of atomic condensate. Wave packet
 dynamics in the absence of modulation (a),and in the presence of modulation 
when $q=3\beta_0$ (b). Other
 parameters are $\acute V=0.36$, $\Delta p=0.1$ and $k^{\hspace{-2.1mm}-}=0.16$}.%
 \label{fig:Spatiotemp}%
 \end{figure}
Quantum revival time in nonlinear resonances versus modulation is shown in
middle column of Fig. \ref{fig:recure}. For robust dynamical recurrences, 
the behavior of quantum revival time is given by equation
(\ref{TdeepRev}). The qualitative behavior of
revival time for strongly driven shallow lattice is different from that of
strongly driven deep lattice Fig. \ref{fig:recure}d, as in the later case,
change in revival time is
almost one order of magnitude larger than the former case, for equal
changes in modulation.  Here, the difference in the revival time
behavior for strongly driven shallow lattice as compared to strongly driven
deep lattice is due to the difference in energy spectrum of undriven system.
In deep lattice, due to small non-linearity more energy bands are influenced by
the external drive and resonance spectrum is similar to that of harmonic
oscillator near the
center of nonlinear resonance. As modulation is
increased more and more energy bands are influenced by external drive.
Fig. \ref{fig:Spatiotemp} shows spatio-temporal evolution of an initially well
localized wave
packet in a lattice potential well inside a resonance 
for $\acute V=16$. 
Fig. \ref{fig:Spatiotemp}a is for the
spatio-temporal dynamics of atomic condensate
in the absence of periodic modulation, while Fig. \ref{fig:Spatiotemp}b presents the case
when external modulation, $q=3\beta_0$.  The quantum revival times in Fig. \ref{fig:Spatiotemp} 
seen numerically are the same as obtained analytically in equation 
(\ref{TdeepRev}).

The behavior of super revival time  versus modulation is shown in right column of 
Fig. \ref{fig:recure}. For robust dynamical
recurrences, equation (\ref{TdeepSup}) gives the time scale for super revivals. 
The super revival time for
robust dynamical recurrences increases with modulation, as shown in the
Fig. \ref{fig:recure}f. Here, qualitative behavior of super revival time is
same for strongly driven shallow lattice and strongly driven  deep lattice but quantitatively 
super revival time increases almost two times faster.
 
\begin{figure}[h]
 \includegraphics[scale=0.20]{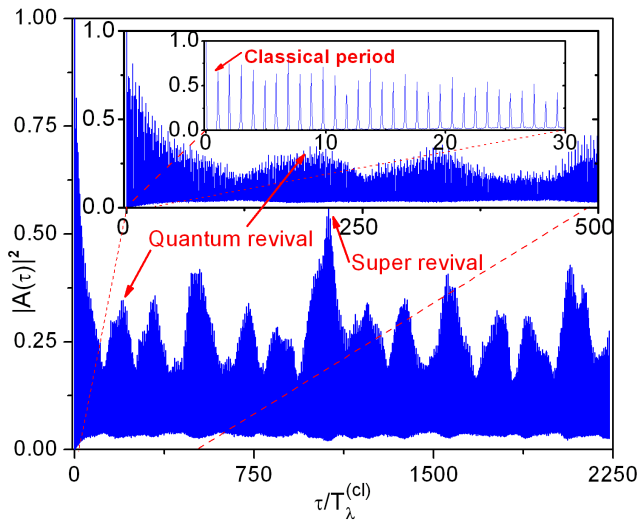}
  \caption{Square of auto-correlation function of Gaussian condensate plotted as a 
function of scaled evolution time. Here,
 $q=0.5\beta_0$, $k^{\hspace{-2.1mm}-}=0.5$, $\acute V=16$, $\Delta z=\Delta
 p=0.5$ and condensate is initially, well localized at the central lattice
 well around the lowest band of undriven crystal. Classical period, revival
 time and super revival time are indicated by arrows.}%
 \label{fig:Auto05}%
 \end{figure}

\begin{figure}[h]
\includegraphics[scale=0.28]{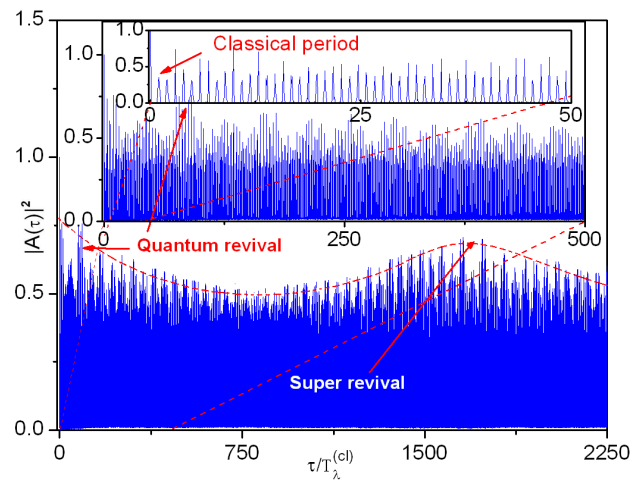}
  \caption{Square of auto-correlation function of Gaussian condensate is plotted versus
scaled time. Here,
 $q=1.5\beta_0$, other parameters and conditions are same as in Fig. \ref{fig:Auto05}}%
 \label{fig:Auto15}%
 \end{figure}

Square of auto-correlation function (${\mid A(t)\mid}^{2}$) for the
minimum uncertainty condensate is plotted as a function of time in
Fig. \ref{fig:Auto05} and Fig. \ref{fig:Auto15} for $q=0.5\beta_0$ and $q=1.5\beta_0$ 
respectively.
Other parameters are $\Delta p=\Delta z=0.5$, 
$k^{\hspace{-2.1mm}-}=0.5$ and $\acute V=16$. In
these figures upper inset is an enlarged view which displays classical
periods, while lower inset of the
figures show quantum revivals. Lower panel shows the existence of super
revivals. The classical period, quantum
revival time and super revival time are indicated by arrows and showing there
characteristics. Numerical results are in good agreement with analytical expressions.

\section{Discussion}
\label{sec:NR}
We presented above a discussion on the occurrence of delicate and robust
dynamical recurrences  of matter wave in optical crystal in
the presence of periodic modulation. 
The delicate dynamics condition, i. e.,  $q\lesssim1$
is satisfied when shallow or deep potential is slightly
perturbed by small external periodic force. Classical periods of 
a condensate initially localized near the center of resonance 
increases with modulation for delicate dynamics. While, quantum revivals and super revivals
decrease with modulation.
Similarly, condition $q\gg1$ can be satisfied for example when shallow or deep potential is
strongly modulated. Classical periods in robust case decrease and super revival 
times increase with increasing modulation. Here, quantum revival time for the case when deep lattice is 
strongly modulated increases with modulation, while, it decreases when shallow 
is strongly modulated.
The difference in the behavior is due to the contrast in energy spectrum of undriven 
crystals. When modulation is 
increased in deep lattice case more and more energy levels are influenced by external 
modulation 
(Fig. \ref{fig:FigPhasespace}e, \ref{fig:FigPhasespace}f), non-linearity 
near the center of resonance decreases and revival time increases.

Parametric dependence of analytical 
results are confirmed by exact  numerical solutions both for delicate and robust dynamical 
recurrences.
Both spatio-temporal (Fig. \ref{fig:Spatiotemp2} and Fig. \ref{fig:Spatiotemp}) and 
temporal (Fig. \ref{fig:Auto05} and Fig. \ref{fig:Auto15}) dynamics show
that the non-linearity of the un-driven system, and the initial conditions on the
excitation contribute to the quantum revival time and super
revival time. 

The suggested theoretical results may be realized in experimental set up of recently performed experiments at Pisa 
\cite{Lignier2007}, where, 
dynamical control of matter wave tunneling is studied in strongly shaken crystal. 
A BECs of about $5\times 10^4$
$^{87}$Rb atoms was evolved in a dipole trap which was realized using two intersecting Gaussian laser
beams at $1030\,\mathrm{nm}$ wavelength and a power of around
$1\,\mathrm{W}$ per beam focused to waists of $50\,\mathrm{\mu
m}$. After obtaining pure condensates trap beams were readjusted to
obtain elongated condensates with the trap frequencies
($80\,\mathrm{Hz}$ in radial and $\approx 20\,\mathrm{Hz}$ in the longitudinal direction). 
Along the axis of one of the dipole
trap beams a one-dimensional optical crystal potential was introduced and the power of the 
lattice beams ramped up in
$50\,\mathrm{ms}$ in order to avoid
excitations of the BEC. The optical crystals used was created using two counter-propagating Gaussian
laser beams ($\lambda_L = 852\,\mathrm{nm}$) with $120\,\mathrm{\mu
m}$ waist and a resulting optical crystal spacing $d_L= \lambda_L /2
= 0.426 \,\mathrm{\mu m}$. The depth $V_o$ of the resulting
periodic potential is measured in units of $E_{\rm r}= \hbar^2
\pi^2 / (2M d_L^2)$. In laboratory, accessible scaled optical
lattice depth $\acute V$ ranges from $1$ to $20$.  

For optical crystal with potential depth $\acute V=2$, and $\acute V=16$, the mean separation of the two lowest bands 
is $\approx3.15kHz$ and $20.784kHz$, respectively.
For the driving frequency, $\omega_m/2\pi$, ranging from  $3kHz-9kHz$, the rescaled Planck's constant 
$k^{\hspace{-2.1mm}-}$ ranges from $0.668$ to $2.066$.
\section{Acknowledgement}
M. A. Thanks HEC Pakistan for financial support through grant
no. 17-1-1(Q.A.U)HEC/Sch/2004/5681.   F. S. thanks HEC Pakistan for partial
financial support under NRP-20-1374.

\end{document}